\documentclass[aps,prl,twocolumn,superscriptaddress]{revtex4}
\usepackage{graphicx}% Include figure files
\usepackage{dcolumn}% Align table columns on decimal point
\usepackage{bm}% bold math
\usepackage{verbatim}
\usepackage{color}
\usepackage[colorlinks]{hyperref}
\usepackage[sort&compress]{natbib}
\input{epsf}
\usepackage[sort&compress]{natbib}
\begin{document}

%\preprint{APS/123-QED}

\title{Observation of Resistively Detected Hole Spin Resonance and Zero-field Pseudo-spin Splitting in Epitaxial Graphene}

\author{Ramesh G. Mani}
\affiliation{Department of Physics and Astronomy, Georgia State
University, Atlanta, GA 30303, U.S.A.}

\author{John Hankinson}
\affiliation {School of Physics, Georgia Institute of Technology,
837 State Street, Atlanta, GA 30332, U.S.A.}

\author{Claire Berger}
\affiliation{School of Physics, Georgia Institute of Technology,
837 State Street, Atlanta, GA 30332, U.S.A.} \affiliation{CNRS,
Institut N\'{e}el, Grenoble, 38042 France}
\author{Walter A. de Heer}
\affiliation{School of Physics, Georgia Institute of Technology,
837 State Street, Atlanta, GA 30332, U.S.A.}
\date{\today}% It is always \today, today,
%\begin{comment}
\begin{abstract}
Electronic carriers in graphene show a high carrier mobility at
room temperature. Thus, this system is widely viewed as a
potential future charge-based high-speed electronic-material to
complement- or replace- silicon. At the same time, the spin
properties of graphene have suggested improved capability for
spin-based electronics or spintronics, and spin-based quantum
computing. As a result, the detection, characterization, and
transport of spin have become topics of interest in graphene. Here
we report a microwave photo-excited transport study of monolayer
and trilayer graphene that reveals an unexpectedly strong
microwave-induced electrical-response and dual microwave-induced
resonances in the dc-resistance. The results suggest the resistive
detection of spin resonance, and provide a measurement of the
g-factor, the spin relaxation time, and the sub-lattice
degeneracy-splitting at zero-magnetic-field.
\end{abstract}
%\end{comment}
%\pacs{72.20.Fr, 73.40.Kp, 72.20.My}% PACS, the Physics and Astronomy
\maketitle

%\section{introduction}

The quantum mechanical spin degree of freedom finds remarkable
applications in the areas of quantum computing (QC) and spin-based
electronics (spintronics).\cite{1,2,3,4,5,6,7,8} For example, in
QC scenarios, particle spin often serves as a quantum bit or
qubit.\cite{1,2,3,4,5,6} In spintronics, the spin serves to endow
electronic devices with new functionality as in the
giant-magneto-resistive (GMR) read-head or the spin-based
transistor.\cite{7,8} Graphene is a novel two-dimensional system
with remarkable properties such as massless Dirac fermions, an
anomalous Berry's Phase, a pseudo-spin (valley-degeneracy) in
addition to spin, and half-integral quantum Hall effect.\cite{
10,11,12,13} Graphene is also an appealing material for
electron-spin QC and spintronics,\cite{1,4,5,6,7,8,14,15,16} due
to the expected weak spin-orbit interaction, and the scarcity of
nuclear spin in natural carbon. Because of QC and spintronics, the
microwave control and electrical detection of spin have become
topics of interest, now in graphene
nanostructures,\cite{1,2,3,4,5,6,7,8,10,11,12,13,14,15,16,17,18}
where the small number of spins limits the utility of traditional
spin resonance.

Here, we report the first observation of resistive detection of
spin resonance in epitaxial graphene,\cite{10,19,21} provide a
measurement of the g-factor and the spin relaxation time, and
determine the pseudo-spin (valley - degeneracy)-splitting at
zero-magnetic-field. Such resistive resonance detection can
potentially serve to directly characterize the spin properties of
Dirac fermions, and also help to determine- and tune- the valley
degeneracy splitting for spin based QC. \cite{16}
\section{Results}
%\subsection{Trilayer graphene}

%\textbf{RESULTS}

\textbf{Trilayer graphene.} Figure 1(a) exhibits the diagonal
resistance, $R_{xx}$, vs. the magnetic field, $B$, for the
trilayer epitaxial graphene specimen, sample 1. The blue curve
obtained at $T = 1.5$K in sample 1 exhibits a cusp in $R_{xx}$
near null magnetic field, i.e., Weak Localization (WL),\cite{20,
22, 23} followed by positive magneto-resistance at $B > 0.2 T$. In
Fig. 1(a), an increase in $T$, to $T = 90K$, results in the red
curve, which includes a positive displacement of the $R_{xx}$ vs.
$B$ trace with respect to the $T=1.5K$ trace, i.e., $dR_{xx}/dT >
0$ at $B = 0$ Tesla, along with the quenching of WL. Since WL
cannot be observed without inter-valley scattering in monolayer or
bilayer graphene,\cite{23} the observed WL is presumed to be an
indicator of a non-zero inter-valley matrix-element.

%\begin{comment}
\begin{figure}[h]
%h=here, t=top, b=bottom, p=separate figure page
\begin{center}
\leavevmode \epsfxsize=2.5 in \epsfbox {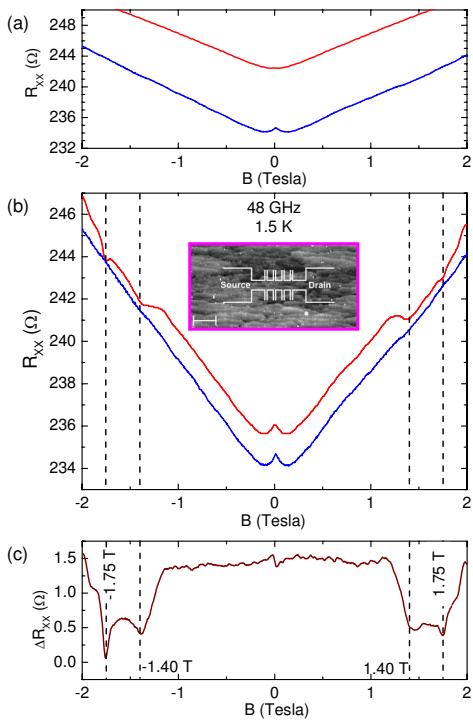}
\end{center}
%\begin{figure}
%\includegraphics{fig1}% Here is how to import EPS art
\caption{\label{fig:epsart} (color online)\textbf{Transport in
trilayer graphene} (a) The diagonal resistance, $R_{xx}$, is shown
vs. the magnetic field, $B$, at temperatures $T = 90$K, shown in
red, and $T = 1.5$ K, shown in blue, for sample 1, in the dark
condition, without microwave excitation. The upward displacement
of the $T = 90$ K curve with respect to the $T = 1.5$ K curve
shows that $R_{xx}$ increases with the temperature, i.e.,
$dR_{xx}/dT \ge 0$. (b) $R_{xx}$ versus $B$ in the absence of
microwave excitation is exhibited in blue, and under constant $F =
48$ GHz microwave excitation at $P = 4$ mW is shown in red, for
sample 1. The photo-excited $R_{xx}$ trace shown in red exhibits a
uniform upward shift with respect to the dark $R_{xx}$ curve shown
in blue for $B < 1$ T. At higher $B$, resonant reductions in the
$R_{xx}$ are observed in the vicinity of $B = \pm 1.4$ T and $B =
\pm 1.75$ T, where the photoexcited $R_{xx}$ approaches the dark
value. Inset: An atomic force microscopy image of the EG/SiC
surface with the device superimposed upon it. The size scale bar
corresponds to 10 $\mu m$. (c) The change in the diagonal
resistance, $\Delta R_{xx}$, between the photo-excited and dark
conditions in panel (b), that is, $\Delta R_{xx} = R_{xx} (4 mW) -
R_{xx} (dark)$, is exhibited versus $B$. Note the valleys in
$\Delta R_{xx}$ in the vicinity of $B = \pm 1.40$ T and $B = \pm
1.75$ T. }\end{figure}
\begin{figure*}[t]
%h=here, t=top, b=bottom, p=separate figure page
\begin{center}
\leavevmode \epsfxsize=6 in \epsfbox {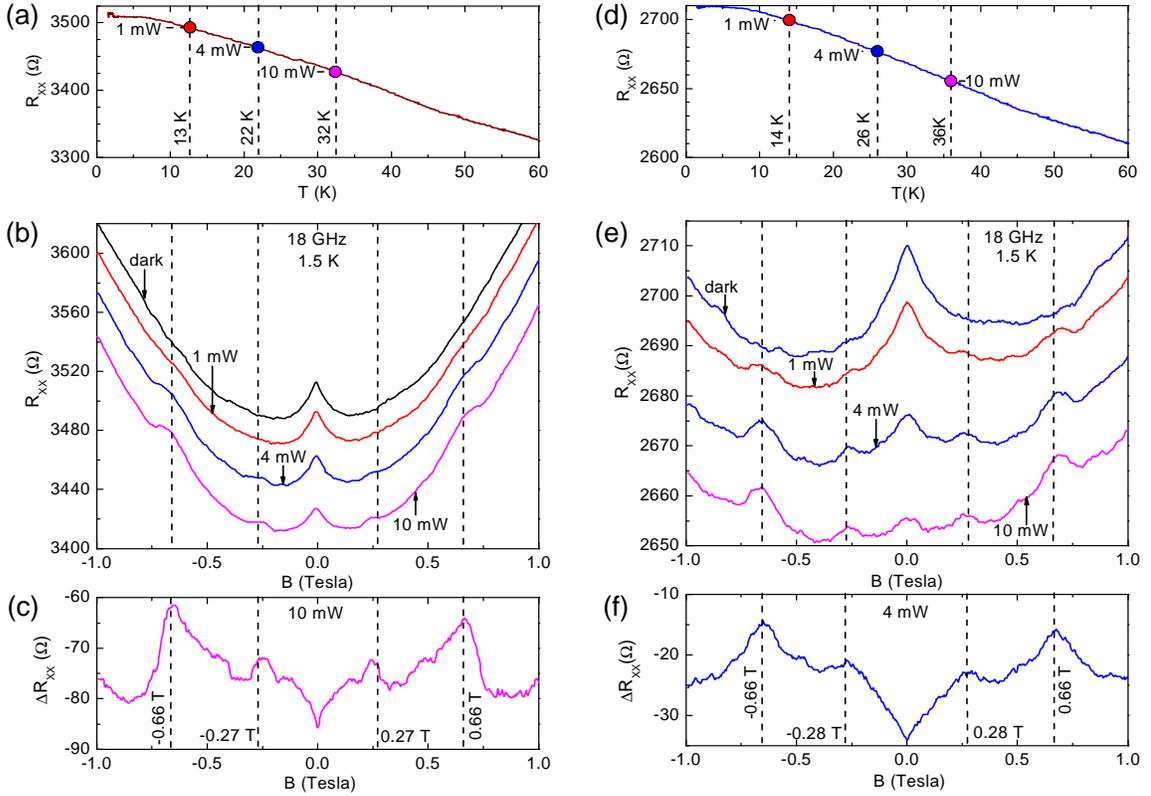}
\end{center}
%\begin{figure}
%\includegraphics{fig1}% Here is how to import EPS art
\caption{\label{fig:epsart} (color online) \textbf{Transport in
single layer graphene} Panels (a)-(c) show some results for sample
2, while figures (d), (e) and (f) show representative data for
sample 3. (a) Sample 2 shows a decrease in the resistance $R_{xx}$
with increasing temperature $T$, i.e., $dR_{xx}/dT \le 0$, in the
absence of a magnetic field, i.e., $B$ = 0 T. (b) The data traces
exhibit a downward shift with increasing microwave power, $P$, at
$F = 18$ GHz, suggestive of microwave induced carrier heating in
the specimen. The $R_{xx}$ at $B = 0$ T obtained from these traces
have been marked as filled circles in panel (a) above, which shows
the T-dependence of $R_{xx}$. Apparently, $P = 10$ mW serves to
increase the carrier temperature up to $T = 32$ K in sample 2.
Microwave induced resonances appear in the vicinity of the dashed
lines with increasing $P$. (c) $\Delta R_{xx} = R_{xx} (10 mW) -
R_{xx}$ (dark) is shown vs. the magnetic field for sample 2. Note
the change in $\Delta R_{xx}$ due to spin-resonance in the
vicinity of the dashed lines at $B = \pm 0.66$ T and $B = \pm
0.27$ T. (d) Sample 3 also exhibits a decrease in the resistance
$R_{xx}$ with increasing temperature $T$, i.e., $dR_{xx}/dT \le
0$, in the absence of a magnetic field, i.e., $B = 0$ T. (e) At $F
= 18$ GHz and $T = 1.5$ K, $R_{xx}$ is shown versus $B$ for sample
3, at several power levels. The $R_{xx}$ at $B = 0$ T observed in
these data have been marked as filled circles in panel (d). (f)
This panel shows $\Delta R_{xx} = R_{xx} (10 mW) - R_{xx} (dark)$
versus the magnetic field for sample 3. Note the change in $\Delta
R_{xx}$ due to spin-resonance in the vicinity of $B = \pm 0.66$ T
and $B = \pm 0.28$ T. }
\end{figure*}
\begin{figure*}[t]
%h=here, t=top, b=bottom, p=separate figure page
%\begin{center}
%\includegraphics[width=3.5in]{fig3pdf}
%\leavevmode \epsfxsize=5.5 in \epsfbox {fig03v2ceps.eps}
\leavevmode \epsfxsize=6 in \epsfbox {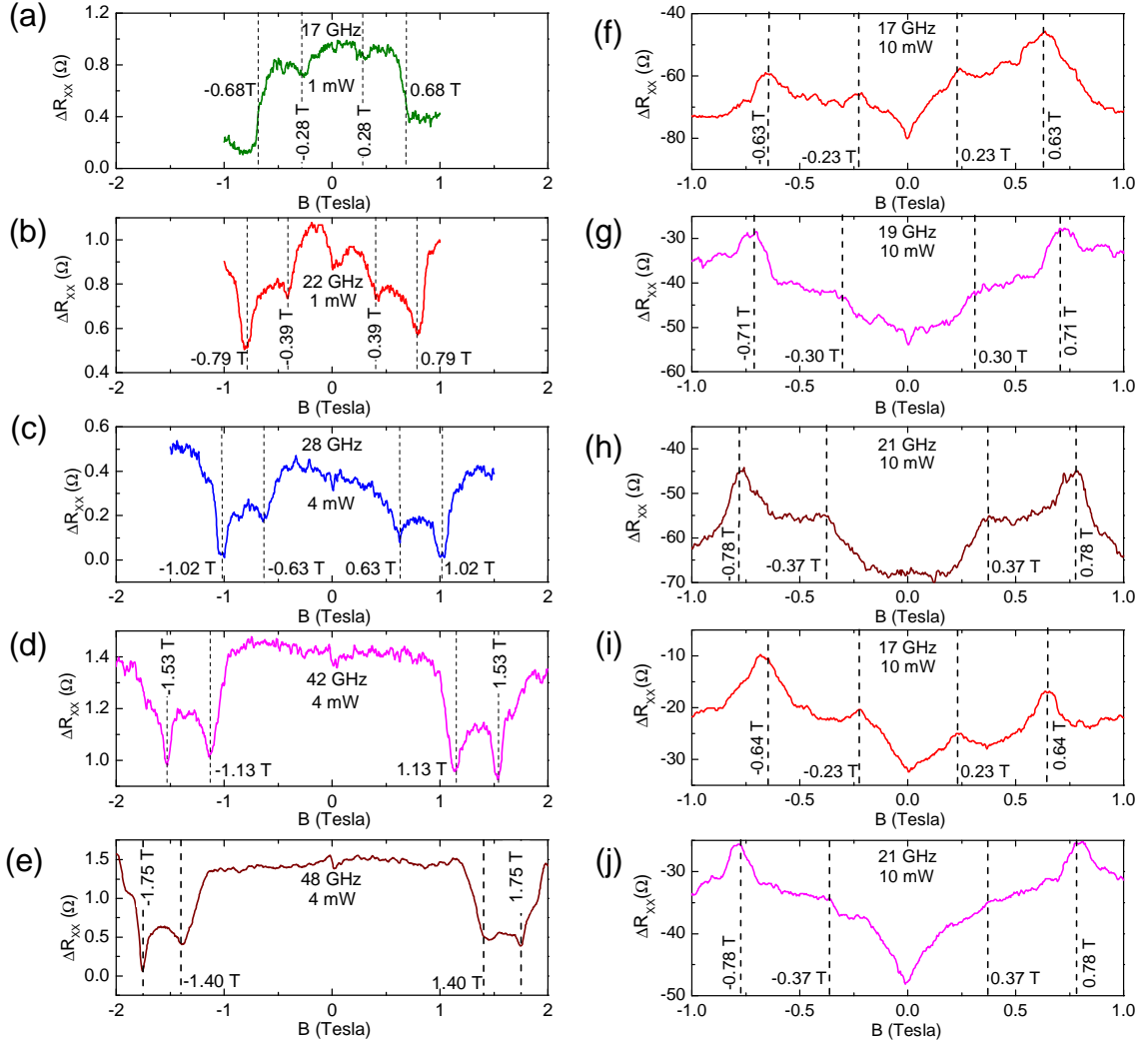}
%\end{center}
%\begin{figure}
%\includegraphics{fig1}% Here is how to import EPS art
\caption{\label{fig:epsart} (color online) Figure 3)
\textbf{Frequency evolution of microwave-induced resonances} (a) -
(e): For sample 1, the photo-induced change in the diagonal
resistance, $\Delta R_{xx}$, is plotted vs. $B$ for (a) 17 GHz,
(b) 22 GHz, (c) 28 GHz, (d) 42 GHz, and (e) 48 GHz. Note the
uniform shift in the resonances, indicated by dashed lines, to
higher $B$ with increasing $F$. Here, the resonances are
characterized by $\Delta R_{xx}$ minima. (f)-(h): For sample 2,
the photo-induced change in the diagonal resistance, $\Delta
R_{xx}$, is plotted vs. $B$ for (f) 17 GHz, (g) 19 GHz, and (h) 21
GHz.  Note the shift in the resonances, indicated by dashed lines,
to higher $B$ with increasing $F$. Here, the resonances are
characterized by resistance maxima. (i)-(j): For sample 3, the
photo-induced change in the diagonal resistance, $\Delta R_{xx}$,
is plotted versus $B$ for (i) 17 GHz and (j) 21 GHz.  Note the
shift in the resonances, indicated by dashed lines, to higher $B$
with increasing $F$.}
\end{figure*}
\begin{figure}[t]
%h=here, t=top, b=bottom, p=separate figure page
\begin{center}
\leavevmode \epsfxsize=3.1 in \epsfbox {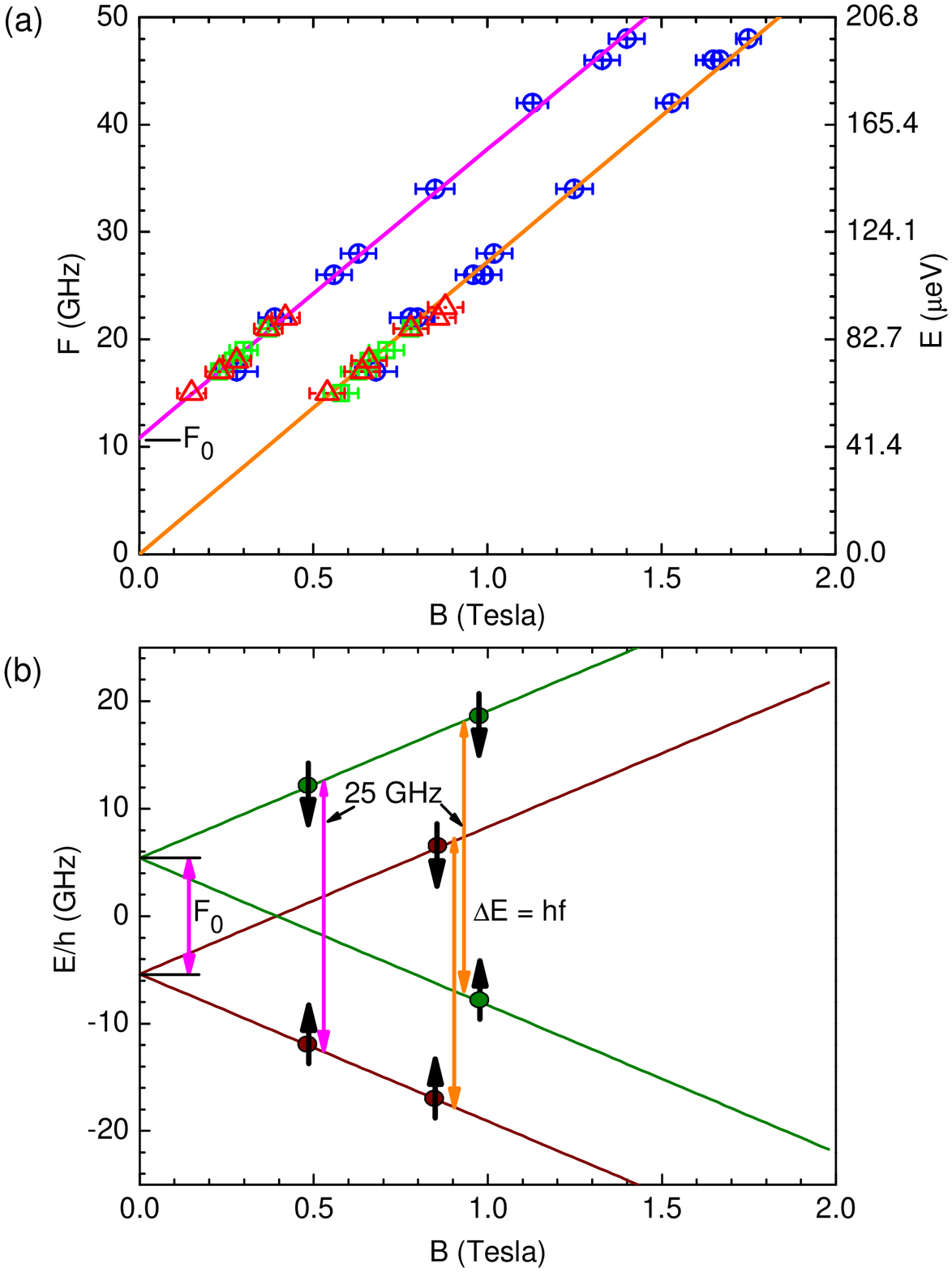}
\end{center}
%\begin{figure}
%\includegraphics{fig1}% Here is how to import EPS art
\caption{\label{fig:epsart} (color online)\textbf{Relation between
the resonance magnetic fields and the microwave frequency or
energy} (a) This plot shows that the high $B$-field resonance
found in Fig. 3 follows a linear fit, indicated by the gold line,
as $F(GHz) = 27.2 B(T)$ with an intercept at the origin. The low
B-field resonance found in Fig. 3, indicated by the magenta line,
follows a linear fit as $F (GHz) = 10.76 + 26.9 B(T)$ with a
non-zero intercept, $F_{0} = 10.8$ GHz. Here, the lines indicate
the fits, while the symbols exhibit the data points. The circles
correspond to the data for sample 1, the squares correspond to the
data for sample 2, and the triangles correspond to the data for
sample 3. (b) The observed experimental results appear to be
consistent with spin-resonance and zero-field pseudo-spin
(valley-degeneracy) splitting enhanced spin resonance. A four-fold
degeneracy is lifted in the absence of a magnetic field to produce
a pair of spin degenerate levels (doublets) separated by $E/h =
F_{0}$. Zeeman splitting then lifts the spin-degeneracy of the
upper and lower doublets.  Microwave photo-excitation induces
spin-flip transitions between the spin-levels of the lower- or
upper doublet, as shown by the gold lines. Such transitions
require vanishing photon energy in the $B \longrightarrow 0$
limit. On the other hand, the transition shown in magenta requires
non-vanishing photon energy in the limit of $B \longrightarrow
0$.}
\end{figure}
%\end{comment}
Fig. 1(b) illustrates the influence of microwave-excitation on
sample 1 at $F = 48$ GHz. Here, for $B < 1$ Tesla,
microwave-excitation produces a positive displacement of the
photo-excited $R_{xx}$ relative to the blue trace obtained in the
absence of photo-excitation, akin to increasing the temperature,
cf. Fig 1(a) and Fig. 1(b). However, at $B > 1$ Tesla, $R_{xx}$
exhibits resistance valleys as the photo-excited curve approaches
the dark curve, similar to reducing the temperature. To highlight
associated resonances, the change in the diagonal resistance,
$\Delta R_{xx} = R_{xx}(4 mW) - R_{xx}(dark)$, is exhibited vs. B
in Fig. 1(c). Fig. 1(c) shows two noteworthy features: a high
magnetic field resonance at $|B| =1.75$ Tesla, and a low magnetic
field feature at $|B| = 1.4$ Tesla. These resonances disappeared
upon increasing the bath temperature to $T
>  5$K.

\textbf{Monolayer graphene.} Figure 2(a)-(c) exhibit the results
for sample 2, while figures 2(d)-(f) show representative data for
sample 3. Both sample 2 and sample 3 are monolayer epitaxial
graphene specimens. The T-dependence of $R_{xx}$ at $B = 0$ Tesla
is shown in Figure 2(a),(d) for samples 2 and 3, respectively.
Unlike sample 1, samples 2 and 3 show a decrease in $R_{xx}(B=0)$
with increasing temperature, i.e., $dR_{xx}/dT \le 0$. Further, as
indicated in Fig. 2(b),(e), microwave irradiation of these
specimens produces a uniform negative displacement in the $R_{xx}$
traces with increasing power. Yet, the effect of microwave
excitation ($dR_{xx}/dP \le 0$ at $B = 0$) is again similar to
heating the specimen ($dR_{xx}/dT \le 0$), cf. Fig. 2(a),(b) or
Fig. 2(d),(e). Thus, the $R_{xx}(B = 0)$ from Fig. 2(b),(e) have
been marked as colored disks in Fig. 2(a),(d). Apparently,
microwave excitation at $P = 10$ mW serves to increase the
carrier-temperature up to $T = 32$ K in sample 2, and up to $T =
36$ K in sample 3. At such higher $P$, the radiation helps to
manifest, in addition, resonant $R_{xx}$ peaks in the vicinity of
the dashed lines of Fig. 2(b),(e), unlike in Fig. 1(b), where
valleys characterize the resonances in $R_{xx}$. Yet, in all three
specimens, the photo-excited $R_{xx}$ moves towards the dark curve
at resonance. Figs. 2 (c),(f) show resonances at nearly the same
$|B|$, at $F = 18$ GHz, in samples 2 and 3.

\textbf{Spin resonance evolution.} Figure 3 illustrates the
frequency-evolution of the $\Delta R_{xx}$ resonances for all
three specimens. Here, Fig. 3(a)-(f) illustrate the results for
sample 1, Fig. 3(f) - (h) exhibit data for sample 2, and Fig. 3(i)
- (j) show some results for sample 3. Note the shift of resonances
to higher $B$ with increasing $F$.

Figure 4(a) presents a plot of the microwave frequency, $F$, vs.
the resonance magnetic fields, $B$, extracted from Fig. 3. Fig.
4(a) shows that the resonance $B$-values for the three specimens
collapse onto two lines: a gold-colored line in Fig. 4(a), which
represents the high $B$-field resonances of Fig. 3, exhibits a
linear increase as $F(GHz) = 27.2 B(T)$, with the
ordinate-intercept at the origin. Another line shown in magenta in
Fig. 4(a), which represents the low-B resonances of Fig. 3,
exhibits a linear increase as $F(GHz) = 10.76 + 26.9 B(T)$, with a
non-zero intercept, $F_{0} = 10.76$ GHz. In such a plot, spin
resonance for an electron with g-factor $g_{e} = 2.0023$ would
follow: $F(GHz) = 28.01 B(T)$. Thus, the observed slopes, $dF/dB =
26.9\pm 0.4$ GHz $T^{-1}$ ($dF/dB = 27.2 \pm  0.2$ GHz $T^{-1}$)
for the low (high) field resonance correspond to spin resonances
with $g_{//} = 1.92\pm 0.028$ ($g_{//} = 1.94\pm 0.014$).

\section{DISCUSSION}

The g-factors measured here are comparable to the g-values
obtained from traditional ESR-studies of graphite, which have
indicated that the g-factor for $B // c$-axis, $g_{//}$, increases
from $2.05$ at $300$ K to $2.15$ at $77$ K, while, at $T =300$ K,
the g-factor for $B \perp c$-axis, $g_{\perp} =
2.003$.\cite{24,25} In graphite, the g-factor depends upon the
orientation of the $B$ field, the temperature, the location of the
Fermi level, and the sign of the charge carriers, with opposite
g-factor shifts, $\Delta g$, from $g_{e}$, for electrons and
holes.\cite{24,25,26} The negative $\Delta g$ and reduced $g_{//}$
observed here relative to $g_{e}$ are consistent with expectation
for holes.

From these data, we also estimate a resonance half-width $\Delta B
\approx 0.05$ Tesla, which corresponds to a spin relaxation time
$\tau_{s} = h/(4\pi \Delta E) = 6 \times 10^{-11}$ s. The spin
relaxation time has been a topic of great interest in graphene.
Spin relaxation \cite{26z} has been experimentally studied in
monolayer and bilayer exfoliated graphene on
$SiO_{2}/Si$\cite{26a, 26b, 26c, 26d, 26e, 26f, 26g, 26h} and,
more recently, on epitaxial graphene \cite{26l, 26m} using spin
valve devices, and possible mechanisms involved in spin relaxation
have been examined by theory.\cite{26i, 26j, 26k} For monolayer
exfoliated graphene, observed spin relaxation times generally fall
in the range of 40 - 150 ps (refs \cite{26b, 26d, 26e, 26h}) while
exfoliated bilayer graphene exhibits spin relaxation times as long
as 2 - 6 ns (refs \cite{26f, 26g}). The observed
shorter-than-expected spin lifetime in exfoliated monolayer
graphene has been attributed to extrinsic mechanisms based on
impurity adatoms,\cite{26i} charged impurities and phonons from
the substrate,\cite{26j} spin orbit coupling due to ripples in
graphene,\cite{26k} and so on. Note that the $\tau_{s}$ reported
here for $C$-face epitaxial graphene is comparable to previous
reports of the spin relaxation time for monolayer exfoliated
graphene on $SiO_{2}/Si$. The spin-diffusion length here is
$\lambda_{s} =(D \tau_{s})^{1/2}= 1.4 \mu m$,\cite{14} while the
Hall bar width, $w = 4 \mu m$. Here, $D$ is the diffusion
constant. In such a situation, edges could be playing a role in
spin relaxation, given that edges in graphene can be magnetically
active,\cite{13} and the electrical contacts include gold, a heavy
element. Thus, there could to be additional avenues for spin
relaxation in the small specimen, in addition to the other
above-mentioned mechanisms.\cite{26i, 26j, 26k} Finally,
inhomogeneities could serve to broaden the resonance linewidth and
help to produce an apparently reduced spin relaxation time.

The observation of similar double resonances in monolayer- and
trilayer- graphene can be viewed as a consequence of rotational
(non-AB) layer stacking in epitaxial graphene, which makes it
possible even for multilayer EG to exhibit the same electronic
properties as isolated graphene.\cite{19} Note also that
sub-lattice or pseudo-spin degeneracy lifting is known to occur at
high-B in graphene.\cite{28,29,30} For example, the progression of
quantum Hall effect from the $R_{xy}$ =  $[4(N + 1/2)]^{-1}$
$h/e^{2}$ sequence,\cite{11,12,27} to observations of
$\sigma_{xy}$ increases in steps of $e^{2}/h$ (ref \cite{28})
reflects the lifting of both the spin- and pseudo-spin-
degeneracy. In addition, a non-linear interaction-enhanced
valley-degeneracy splitting has been reported from a scanning
tunneling spectroscopy study.\cite{30} Finally, the manifestation
of weak localization, which is observable in Fig. 1 and 2, is an
indicator of inter-valley coupling in these specimens.\cite{23}
Since sub-lattice degeneracy splitting is not unexpected due to
the above, the observed $F_{0} = 10.76$ GHz is attributed to a
zero-magnetic-field pseudo-spin (sub-lattice degeneracy) splitting
of $\Delta_{0} = h F_{0} = 44.4 \mu$ eV.

A provisional interpretation of the F vs. B plot of Fig. 4(a) is
provided in Fig. 4(b). Chiral eigenstates and linear
energy-wavevector dispersion characterize carriers in graphene.
The application of a B-field nominally produces fourfold, valley-
and spin- degenerate Landau-levels characterized by $E_{N} = \pm
v_{F} (2e \hbar BN)^{1/2} $, where $N = 0,  1, 2, \cdots$, $e$ is
the electron charge, $v_{F}$ is the Fermi velocity, and $\hbar$ is
the reduced Planck's constant.

We imagine the four-fold degeneracy being lifted by $hF_{0}$ even
at $B=0$, to produce energy doublets as $E_{N'} = E_{N} \pm
hF_{0}/2$. Then, owing to the Zeeman effect, associated Landau
levels exhibit a further splitting of the spin-degeneracy as
$E_{N''} = E_{N'} \pm g\mu B/2$. Observed microwave-induced
transitions occur within the highest occupied Landau level in the
vicinity of the Fermi level. As $E_{N} >> hF_{0}/2$ and $g \mu
B/2$, we remove the $E_{N}$ term and plot $E/h = (E_{N''} -
E_{N})/h$ in Fig. 4(b).

Here, microwave photo-excitation induces spin-flip transitions,
shown in gold, of unpaired carriers between the spin-levels of the
lower or the upper doublet. Such transitions require vanishing
photon energy in the limit of vanishing $B$. In contrast, a
transition between the lower spin ("up") level of the lower
doublet and the higher spin ("down") level of the upper doublet
requires additional energy $hF_{0}$, and such a transition, shown
in magenta, exhibits non-vanishing photon energy in the $B
\longrightarrow 0$ limit. Thus, the $F$ vs. $B$ plot appears
consistent with spin resonance and a zero-field pseudo-spin
(valley-degeneracy) splitting enhanced spin resonance.

In summary, we have realized the resistive detection of spin
resonance in EG, provided a measurement of the g-factor and the
spin relaxation time, and identified- and measured- a pseudo-spin
(valley degeneracy)-splitting in the absence of a magnetic field.
Such resistive resonance detection can potentially serve to
directly characterize the spin properties of Dirac fermions, and
also help to determine- and tune- the valley degeneracy splitting
for spin based QC.

\section{METHODS}

\textbf{Graphene Samples.} Epitaxial graphene (EG) was realized by
the thermal decomposition of insulating 4H silicon carbide
(SiC).\cite{19} The EG specimens were characterized by
ellipsometry and the extracted layer thickness was converted to
the number of layers at the rate of 0.335 nm/layer. The c-face of
the EG/SiC chip was processed by e-beam lithography into
micron-sized Hall bars with Pd/Au contacts. Measurements are
reported here for three Hall bar specimens labeled 1, 2, and 3.
Sample 1 is nominally trilayer graphene, while samples 2 and 3 are
monolayer graphene.  The samples are p-type with a hole
concentration, $p \approx 10^{13} cm^{-2}$, and a carrier mobility
$\mu \approx 10^{3} cm^{2}/Vs$.

\textbf{Measurement configuration.} Typically, an epitaxial
graphene Hall bar specimen was mounted at the end of a long
straight section of WR-62 rectangular microwave waveguide. The
waveguide with sample was inserted into the bore of a
superconducting solenoid, immersed in pumped liquid Helium, and
irradiated with microwaves over the frequency range $10 \le F \le
50$ GHz, at a source-power $0.1 \le P \le 10$ mW, as in the usual
microwave-irradiated transport experiment.\cite{31}  Here, the
applied external magnetic field was oriented along the solenoid
and waveguide axis as a probe coupled antenna launcher excited the
Transverse Electric (TE-10) mode in the waveguide.  Thus, the
microwave electric field was oriented perpendicular to the applied
external magnetic field. The microwave magnetic field lines formed
closed loops, with components in the transverse and axial
directions of the waveguide.


\newpage

\section{ACKNOWLEDGEMENTS}

The basic research at Georgia State University is primarily
supported by the U.S. Department of Energy, Office of Basic Energy
Sciences, Material Sciences and Engineering Division under
DE-SC0001762 through A. Schwartz. Additional support is provided
by D. Woolard and the ARO under W911NF-07-01-015. Work at the
Georgia Institute of Technology is supported by the W. M. Keck
Foundation, AFOSR, and the NSF under grant DMR-0820382. We
acknowledge discussions with V. Apalkov, M. Kindermann, and S.
Adam.

\section{AUTHOR CONTRIBUTIONS}

The microwave transport experiments and the manuscript are due to
R.G.M. The graphene samples were grown and processed by J.H., C.
B., and W.A.d.H. All authors contributed to the discussion.

\section{ADDITIONAL INFORMATION}

Competing financial interests: The authors declare no competing
financial interests.

Correspondence and requests for materials should be addressed to
R.G.M (email: mani.rg@gmail.com)

\newpage
\pagebreak
%\begin{comment}
\begin{comment}
%%%%%%%%%%%%%%%%%%%%%%%%%%%%%%%%%%%%%%%%
\begin{figure*}[p]
%h=here, t=top, b=bottom, p=separate figure page
\begin{center}
\leavevmode \epsfxsize=4.25 in \epsfbox {fig1eps.eps}
\end{center}
%\begin{figure}
%\includegraphics{fig1}% Here is how to import EPS art
\caption{\label{Fig. 1}}
\end{figure*}

\pagebreak
\begin{figure*}[p]
%h=here, t=top, b=bottom, p=separate figure page
\begin{center}
\leavevmode \epsfxsize=5.25 in \epsfbox {fig2eps.eps}
\end{center}
%\begin{figure}
%\includegraphics{fig2}% Here is how to import EPS art
\caption{\label{Fig. 2}}
\end{figure*}

\pagebreak
\begin{figure*}[p]
%h=here, t=top, b=bottom, p=separate figure page
\begin{center}
\leavevmode \epsfxsize=6.25 in \epsfbox {fig3eps.eps}
\end{center}
%\begin{figure}
%\includegraphics{fig2}% Here is how to import EPS art
\caption{\label{Fig. 3}}
\end{figure*}

\pagebreak
\begin{figure*}[p]
%h=here, t=top, b=bottom, p=separate figure page
\begin{center}
\leavevmode \epsfxsize=4.25 in \epsfbox {fig4eps.eps}
\end{center}
%\begin{figure}
%\includegraphics{fig2}% Here is how to import EPS art
\caption{\label{Fig. 4}}
\end{figure*}
\end{comment}
%%%%%%%%%%%%%%%%%%%%%%%%%%%%%%%%%%%%%%%%
\end{document}